\documentclass[useAMS,usenatbib,usegraphicx]{mn2e}
\usepackage{hyperref}

\def \mbh {M_{\rmn{BH}}}
\def \sigc {\sigma_{\rmn{GC}}}
\def \kms {\rmn{km \ s^{-1}}}

\title[$\mbh-\sigma$ relation between SMBHs and the velocity dispersion of globular cluster systems]
{$\mbh-\sigma$ relation between SMBHs and the velocity dispersion of globular cluster systems}

\author[Raphael Sadoun and Jacques Colin]
{Raphael Sadoun$^{1}$ and %\thanks{E-mail: colin@iap.fr} 
 Jacques Colin$^{1}$ \\
$^{1}$ UPMC, CNRS, Institut d'Astrophysique de Paris, 98 bis Bd. Arago, Paris 75014, FRANCE
}

\begin{document}

\date{\today}

\pagerange{\pageref{firstpage}--\pageref{lastpage}} \pubyear{2012}

\maketitle

\label{firstpage}

\begin{abstract}
We find evidence that the mass $\mbh$ of central supermassive black 
holes (SMBHs) correlates with the velocity
dispersion $\sigc$ of globular cluster systems of their host galaxies. 
This extends the well-known $\mbh-\sigma_{sph}$ relation between 
black hole mass and velocity dispersion of the host spheroidal component. 
We compile published measurements of both $\mbh$ and $\sigc$ for a sample of 13 systems and 
find the relation $\log(\mbh)=\alpha+\beta \log(\sigc/200)$ with 
$\alpha = 8.63 \pm 0.09$ and $\beta = 3.76 \pm 0.52$. We also consider  
blue (metal-poor) and red (metal-rich) globular clusters sub-populations separately
and obtain a surprisingly tight correlation using only the velocity dispersion $\sigc^{\rmn{red}}$ 
of the red clusters with $\alpha = 8.73 \pm 0.09$ and $\beta = 3.84 \pm 0.52$ and 
an intrinsic scatter $\varepsilon_0 = 0.22$ dex compared to $\varepsilon_0 = 0.27$ dex for the 
$\mbh-\sigma_{sph}$ of our sample. We use this $\mbh-\sigc^{\rmn{red}}$ relation to estimate the
central black hole mass in five galaxies for which $\sigc^{\rmn{red}}$ is measured.
\end{abstract}

\begin{keywords}
black hole physics -- galaxies: evolution -- galaxies: nuclei -- galaxies: star clusters: general -- 
galaxies: fundamental parameters -- globular clusters: general
\end{keywords}

\section{Introduction}

It has been known for more than a decade that supermassive 
black holes (SMBHs) reside at the centre of many galaxies \citep{Kormendy1995} 
and that their mass correlates with various properties of their host galaxies, for example with the 
spheroid luminosity \citep{Magorrian1998}, the spheroid velocity 
dispersion $\sigma_{sph}$ (\citealt{Ferrarese2000} and \citealt{Gebhardt2000}, 
see also \citealt{Gultekin2009}, hereafter G09), the spheroid mass
\citep{Magorrian1998,Marconi2003} and the kinetic energy of random motions \citep{Feoli2009}. 
Recently, extension of these correlations to barred galaxies and pseudo-bulges
\citep{Hu2008,Graham2008a,Graham2008b,Graham2011} as well as galaxies hosting nuclear star clusters 
\citep{Ferrarese2006,Graham2012} has also been proposed. In addition, a new  
correlation between the mass of SMBHs and the observed total number of globular clusters of their host galaxies 
has been discovered (\citealt{Burkert2010}, hereafter BT10, \citealt{Harris2011}, see also \citealt*{Snyder2011}). 
The physics underlying these observations is not fully understood, but it is widely believed that feedback processes
are responsible for such correlations \citep{Silk1998}. It is not known whether these 
relations evolve with time or are primordial and have been set at the time of 
formation of the SMBHs and galaxies (\citealt*{DiMatteo2005}, \citealt{Cox2006}, 
\citealt{Hopkins2007}, \citealt*{Hopkins2009}, \citealt{Volonteri2009}).
Furthermore, the physical scale at which the feedback is effective is still not clear.
Studies emphasising that black hole mass is coupled to the dark matter (DM)
potential rather than the spheroid \citep*{Ferrarese2002,Booth2010,Volonteri2011} seem to indicate
that feedback processes operate on large scales even though \citet{Kormendy2011} argue that 
there is no direct relation between DM halos and central black holes. 
Better and more data, extending to higher redshifts, shall shed more light on these questions.

In this paper we provide evidence that the classical $\mbh-\sigma_{sph}$ relation between
black hole mass and spheroid velocity dispersion is also working with the velocity dispersion of 
globular cluster (GC) systems. We use the published data for the velocity dispersion $\sigc$ of the GC systems
and the estimated black hole mass in 13 galaxies. We then study separately
the two sub-populations : blue, metal-poor GCs and red, metal-rich and younger GCs which are generally 
closer to the centre of the galaxy. We show that the correlation degrades from the red to the blue GCs sub-population.
    
This paper is organised as follows. In Section \ref{section:data}, we present the data. 
The correlation between the mass of SMBHs and the velocity dispersions of the GC systems
of their host galaxies is presented in Section \ref{section:m-sigma}. The red and blue 
sub-populations are studied separately in Section \ref{section:m-sigma sub}. 
In Section \ref{section:prediction}, we estimate the black hole mass for 5 galaxies 
and conclude in Section \ref{section:conclusion}.

%%%%%%%%%%%%%%%%%%%%%%%%%%%%%%%%%%%%%%%%%%%%%%%%%%%%%%%%%%%%

%%%%%%%%%%%%%%%%%%%%%%%%%%%%%%%%%%%%
\section{The data} 
%%%%%%%%%%%%%%%%%%%%%%%%%%%%%%%%%%%%

\label{section:data}
We have found 19 galaxies for which measurements of the line-of-sight velocity 
dispersion of the GC system are known. Thirteen of these are accompanied by 
the estimate of the mass of their central black hole. These galaxies are presented 
in Table \ref{table:data}. To our knowledge, the mass of the central black holes 
have not yet been determined for the six other galaxies for which we have the 
velocity dispersion of the GC system : NGC 1407 \citep{Romanowsky2009}, 
NGC 3923 \citep{Norris2012}, NGC 4494 \citep{Foster2011}, 
NGC 4636 \citep{Lee2010}, LMC \citep{Freeman1993} and M33 \citep{Schommer1991,Chandar2002}.
   
%%%%%%%%%%%%%%%%%%%%%%%%%%%%%%%%%%%%%%%%%%%%%%%%%%%%%%%%%%%%%%%%%%%%
\subsection{Velocity dispersion of globular cluster systems}
%%%%%%%%%%%%%%%%%%%%%%%%%%%%%%%%%%%%%%%%%%%%%%%%%%%%%%%%%%%%%%%%%%%%
\label{section:m-sigma}
Galaxies for which we found measurements of $\sigc$ as well as 
$\sigc^{\rmn{red}}$ and $\sigc^{\rmn{blue}}$ are listed in Table \ref{table:data} 
along with the corresponding references. Whenever available, we use the 
value of $\sigc$ corrected for the global rotation of the GC system. 
Comments on individual systems are given below : 

\begin{itemize}

\item{\emph{NGC 1399} -}
We use the results of \citealt{Richtler2004} which give $\sigc$ 
for all GCs and for both red and blue sub-populations. 
\citet{Schuberth2010} give detailed results for different sub-samples 
and different distances which are in general not too different 
from \citet{Richtler2004}.

\item{\emph{NGC 3031 (M81)} -}
Results have been obtained separately by \citet{Perelmuter1995}, 
\citet{Schroder2002} and \citet{Nantais2010}. We use the values 
given by \citet{Nantais2010} which are consistent with the other groups. 
Measurement uncertainties in $\sigc$, $\sigc^{\rmn red}$ and $\sigc^{\rmn blue}$ were 
provided by \citet{Nantais2012}.

\item{\emph{NGC 4594 (M104)} -}
The value of $\sigc$ is taken from \citet{Bridges2007}. They also measured $\sigc^{\rmn{red}}$ 
and $\sigc^{\rmn{blue}}$ but do not give the corresponding error bars. Consequently, 
we do not include M104 in our best-fit estimate when treating red and blue GCs
separately.

\item{\emph{MW} -}
\citet{Zinn1996} gives $\sigc$ for three groups of GCs lying approximately 
around the direction of the galactic centre. The corresponding line-of-sights are therefore 
nearly parallels and we can treat the measured radial velocities as projection on a 
single line-of-sight similarly to what is done for external galaxies. 
The three groups are metal-poor (blue), red clusters lying between 2.7 to 6 kpc 
from the galactic centre, and very metal-rich disc clusters. We choose the dispersion 
of the red clusters ($\sigc^{\rmn{red}} = 61 \pm 10 \ \kms$) to compare with the 
dispersion of red GCs in others galaxies. We note that a similar value was found by \citet{Harris1999}
for metal-rich clusters lying in the range $4-9 \ \rmn{kpc}$. Concerning the metal-poor clusters, 
we retain the value $\sigc^{\rmn{blue}} = 120 \pm 14 \ \kms$ given by \citet{Harris1999}. However, 
we point out that comparison with external galaxies is not straightforward in this case since 
we can no longer consider the projection of cluster velocities on a single line-of-sight.

\end{itemize}

%%%%%%%%%%%%%%%%%%%%%%%%%%%%%%%%%%%%%%%%%%%%%%%%%%%%%%%%%%%%
\subsection{Mass of central black holes}
%%%%%%%%%%%%%%%%%%%%%%%%%%%%%%%%%%%%%%%%%%%%%%%%%%%%%%%%%%%%%

The mass of central black holes in 9 out of the 13 galaxies in our sample, 
namely NGC 224, NGC 1399, NGC 3031, NGC 3379, NGC 4486, NGC 4594, NGC 4649, NGC 5128, 
NGC 7457 and MW have been studied and analysed by G09. To avoid systematic errors and for general 
consistency, we have decided to use these results although we are aware that other values 
also exist in the literature. For NGC 1399 and NGC 5128, G09 give two possible values for the mass. 
We follow their procedure and include both values with a weight of $1/2$ when performing the linear fit, 
a method also used by BT10. For the Milky Way, we use $\mbh = 4.3 \ \times 10^6 M_{\sun}$ from 
\citet{Gillessen2009} (G09 give $\mbh = 4.1 \ \times 10^6 M_{\sun}$).

For NGC 4636 only two upper limits for $\mbh$ are given 
in \citet{Beifiori2009}. We do not incorporate these values in the determination 
of the parameters of the relation. However, we still include a posteriori NGC 4636
in all of the presented figures by taking the mean of the values found in 
\citet{Beifiori2009} as an upper limit for $\mbh$.

%%%%%%%%%%%%%%%%%%%%%%%%%%%%%%%%%  TABLE %%%%%%%%%%%%%%%%%%%%%%%%%%%%%%%%%%%%%%%%%

\begin{table*}
\centering
\begin{minipage}{130mm}
\caption{
  Sample of galaxies with measured line-of-sight velocities of globular 
  clusters and mass of the central black hole.
}
\begin{tabular}{@{}lcccclllc@{}}

\hline
Galaxy &             & Type & $\mbh$      & Ref. & $\sigc$            & $\sigc^{\rm blue}$           & $\sigc^{\rm red}$  & Ref. \\
       &             &      & ($M_{\sun}$) &      & ($\rm km\ s^{-1}$) & ($\rm km\ s^{-1}$)          & ($\rm km\ s^{-1}$)          &      \\
\hline
NGC 224  & M31       & Sb   & $1.5^{+0.9}_{-0.3} \ \times 10^8$  & 1 & $134^{+5}_{-5}$   & $129^{+8}_{-6}$   & $121^{+9}_{-10}$   & 6 \\
NGC 524  &           & S0   & $8.3^{+2.7}_{-1.3} \ \times 10^8$  & 9 & $186^{+29}_{-29}$ & $197^{+39}_{-39}$  & $169^{+43}_{-43}$  & 5  \\
NGC 1316 & Fornax A  & SAB  & $1.5^{+0.75}_{-0.8} \times 10^8$   & 2 & $202^{+33}_{-33}$ &                   &                    & 7 \\
NGC 1399 &           & E1   & $1.3^{+0.5}_{-0.66} \times 10^9$   & 1 &  $274^{+9}_{-9}$  & $291^{+14}_{-14}$ & $255^{+13}_{-13}$   & 8 \\
         &           &      & $5.1^{+0.7}_{-0.7} \ \times 10^8$  &   &                  &                  &                    &     \\ 
NGC 3031 & M81       & Sb   & $8.0^{+2.0}_{-1.1} \ \times 10^7$  & 1 & $128^{+9}_{-9}$  & $141^{+15}_{-15}$  & $125^{+11}_{-11}$  & 10 \\
NGC 3379 &           & E0   & $1.2^{+0.8}_{-0.58} \times 10^8$   & 1 & $175^{+24}_{-22}$  &                  &                   & 11 \\
NGC 4472 & M49       & E4   & $1.5^{+0.2}_{-0.2} \ \times 10^9$  & 4 & $312^{+27}_{-8}$  & $342^{+33}_{-18}$ & $265^{+34}_{-13}$  & 12 \\
NGC 4486 & M87       & E1   & $3.6^{+1.0}_{-1.0} \ \times 10^9$  & 1 & $320^{+11}_{-11}$ & $335^{+15}_{-15}$ & $295^{+23}_{-23}$  & 13 \\
NGC 4594 & M104      & Sa   & $5.7^{+4.4}_{-4.0} \ \times 10^8$  & 1 & $204^{+16}_{-16}$ & $203$            & $207$             & 14 \\
NGC 4649 & M60       & E2   & $2.1^{+0.5}_{-0.6} \ \times 10^9$  & 1 & $217^{+14}_{-16}$ & $207^{+15}_{-19}$ & $240^{+20}_{-34}$  & 16 \\
NGC 5128 & Cen A     & S0/E & $3.0^{+0.4}_{-0.2} \ \times 10^8$  & 1 & $150^{+2}_{-2}$   & $149^{+4}_{-4}$   & $156^{+4}_{-4}$    & 17 \\
         &           &      & $7.0^{+1.3}_{-3.8} \ \times 10^7$  &   &                  &                   &                   &    \\
NGC 7457 &           & S0   & $4.1^{+1.2}_{-1.7} \ \times 10^6$  & 1 & $69^{+12}_{-12}$  &                   &                   & 3  \\
MW       &           & Sbc  & $4.3^{+0.4}_{-0.4} \ \times 10^6$  & 18 &                  &  $120^{+14}_{-14}$ & $61^{+10}_{-10}$   & 15 \\
\hline

\end{tabular}
\textbf{REFERENCES.} (1) \citealt{Gultekin2009}; (2) \citealt{Nowak2008}; (3) \citealt*{Chomiuk2008}; 
(4)\citealt{Shen2012}; (5)  \citealt{Beasley2004}; (6) \citealt{Lee2008}; 
(7) \citealt{Goudfrooij2001}; (8) \citealt{Richtler2004}; (9) \citealt{Krajnovic2009}; 
(10) \citealt{Nantais2010}; (11) \citealt{Bergond2006}; (12) \citealt{Cote2003}; 
(13) \citealt{Strader2011}; (14) \citealt{Bridges2007}; (15) \citealt{Zinn1996}; 
(16) \citealt{Hwang2008}; (17) \citealt{Woodley2010}; (18) \citealt{Gillessen2009}
\label{table:data}
\end{minipage}
\end{table*}

\section{The $\mbh-\sigc$ relation}
%%%%%%%%%%%%%%%%%%%%%%%%%%%%%%%%%%%%%%%%%%%%%%%%%%%%%%%%%%%%
\label{section:m-sigma}

\begin{figure}
  \includegraphics[width=\linewidth]{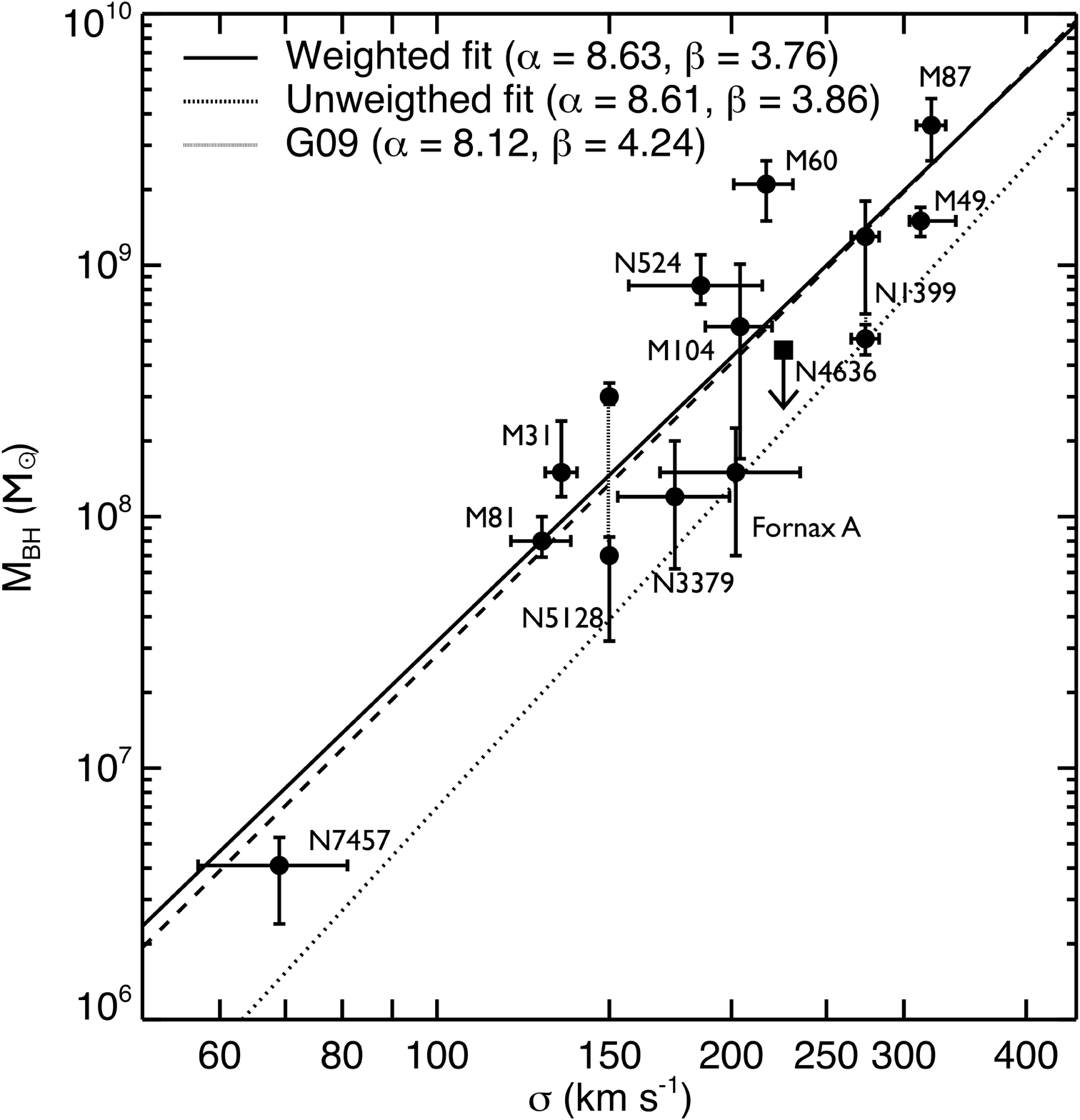}
  \caption{
    Mass $\mbh$ as a function of $\sigc$ for the 12 galaxies in our sample.
    Circles indicate galaxies included in the fitting procedure. NGC 4636, for which only 
    an upper limit of $\mbh$ is available, is shown as a square. The solid black line is the best-fit 
    relation obtained using the weighted $\chi^2$-minimisation procedure of \citet{Tremaine2002}. 
    The dashed line is the best-fit relation found using an unweighted least-squares method 
    without taking into account error bars. The dotted line is the $M-\sigma$ relation of G09.
  }
  \label{figure:m-sigma total}
\end{figure}

In Figure \ref{figure:m-sigma total} we plot the central black hole mass $\mbh$ versus 
the velocity dispersion $\sigc$ of the GC system with the error bars (see Table \ref{table:data}). 
To facilitate comparison with the previous works of G09, we use the same presentation. 
We assume a relation of the form $\rmn{log} \left(\mbh/M_{\sun}\right) = \alpha + 
\beta \  \rmn{log} \left(\sigc/200 \ \kms \right)$.
The parameters of the relation are calculated by the $\chi^2$-minimisation technique of \citet{Tremaine2002}
using the \textsc{IDL} MPFITEXY routine \citep*{Williams2010} which includes error bars in 
both $\mbh$ and $\sigc$ (weighted fit) and allows the determination 
of the intrinsic scatter $\varepsilon_0$ in $\mbh$ at fixed $\sigc$. The MPFITEXY
routine depends on the MPFIT package \citep{Markwardt2009}.We also carry out a
standard linear least-squares fit without taking into account error bars (unweighted fit)
for comparison. We obtain $\alpha = 8.63 \pm 0.09$ and $\beta = 3.76 \pm 0.52$ for the weighted fit (table \ref{tab:params}). 
For the full sample used by G09, these values are $\alpha = 8.12\pm 0.08$ and  $\beta= 4.24\pm 0.41$.
The slope of the relation between $\mbh$ and $\sigc$ is consistent with the one obtained by G09. 
However, for the same $\mbh$ the velocity dispersion of the GC system is systematically smaller than that obtained 
for the spheroid. The intrinsic scatter we found is $\varepsilon_0 = 0.27$. As a comparison, 
we have examined the usual $\mbh-\sigma_{sph}$ relation using the galaxies in our sample
with values of $\sigma_{sph}$ taken from G09 and obtained a similar value of $\varepsilon_0 = 0.27$ with 
the same fitting procedure. Thus, it seems that, for the limited number of galaxies in our sample, 
$\mbh$ correlates equally well with either $\sigma_{sph}$ or $\sigc$.

%%%%%%%%%%%%%%%%%%%%%%%%%%%%%%%%%%%%%%%%%%%%%%%%%%%%%%%%%%%%%%%%%%%%
\section{Relation for blue and red globular cluster systems}
%%%%%%%%%%%%%%%%%%%%%%%%%%%%%%%%%%%%%%%%%%%%%%%%%%%%%%%%%%%%%%%%%%%%

\begin{table}
\centering
\caption{
  Values of the parameter of the relation $\rmn{log}\left(\mbh/M_{\sun}\right) = \alpha 
  + \beta \rmn{log}\left(\sigma/200 \ \kms\right)$ for the full GC system as well 
  as for the red and blue sub-populations.
}
\begin{tabular}{@{}lccc@{}}
\hline
             &     $\alpha$      &     $\beta$      & $\varepsilon_0$ \\
\hline
Full sample  &  $8.63 \pm 0.09$  &  $3.76 \pm 0.52$ & $0.27$ \\
Red  GCs     &  $8.73 \pm 0.09$  &  $3.84 \pm 0.52$ & $0.22$ \\
Blue GCs     &  $8.66 \pm 0.12$  &  $3.00 \pm 0.72$ & $0.33$ \\
\hline
\label{tab:params}
\end{tabular}
\end{table}

\begin{figure}
  \includegraphics[width=\linewidth]{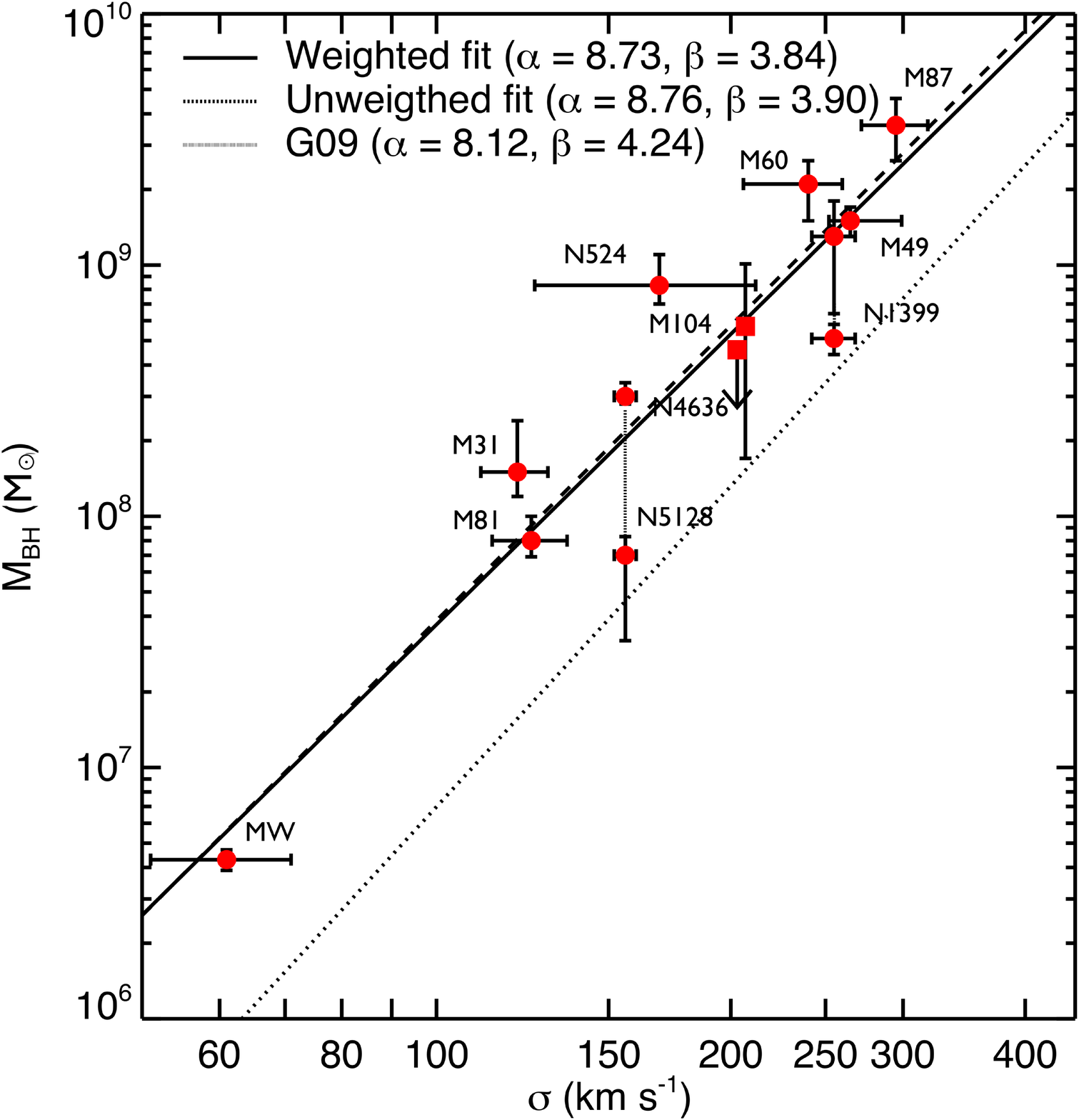}
  \caption{
    Mass $\mbh$ versus velocity dispersion $\sigc^{\rmn{red}}$ of the red, metal-rich
    globular cluster system for galaxies in our sample. As in Figure \ref{figure:m-sigma total}, circles
    are data points included in the fitting procedure whereas squares are systems which do not contribute
    to the best-fit relation, namely NGC4594 and NGC4636. 
    The lines have the same meaning as in Figure \ref{figure:m-sigma total}.
  }
  \label{figure:m-sigma red}
\end{figure}

\begin{figure}
  \includegraphics[width=\linewidth]{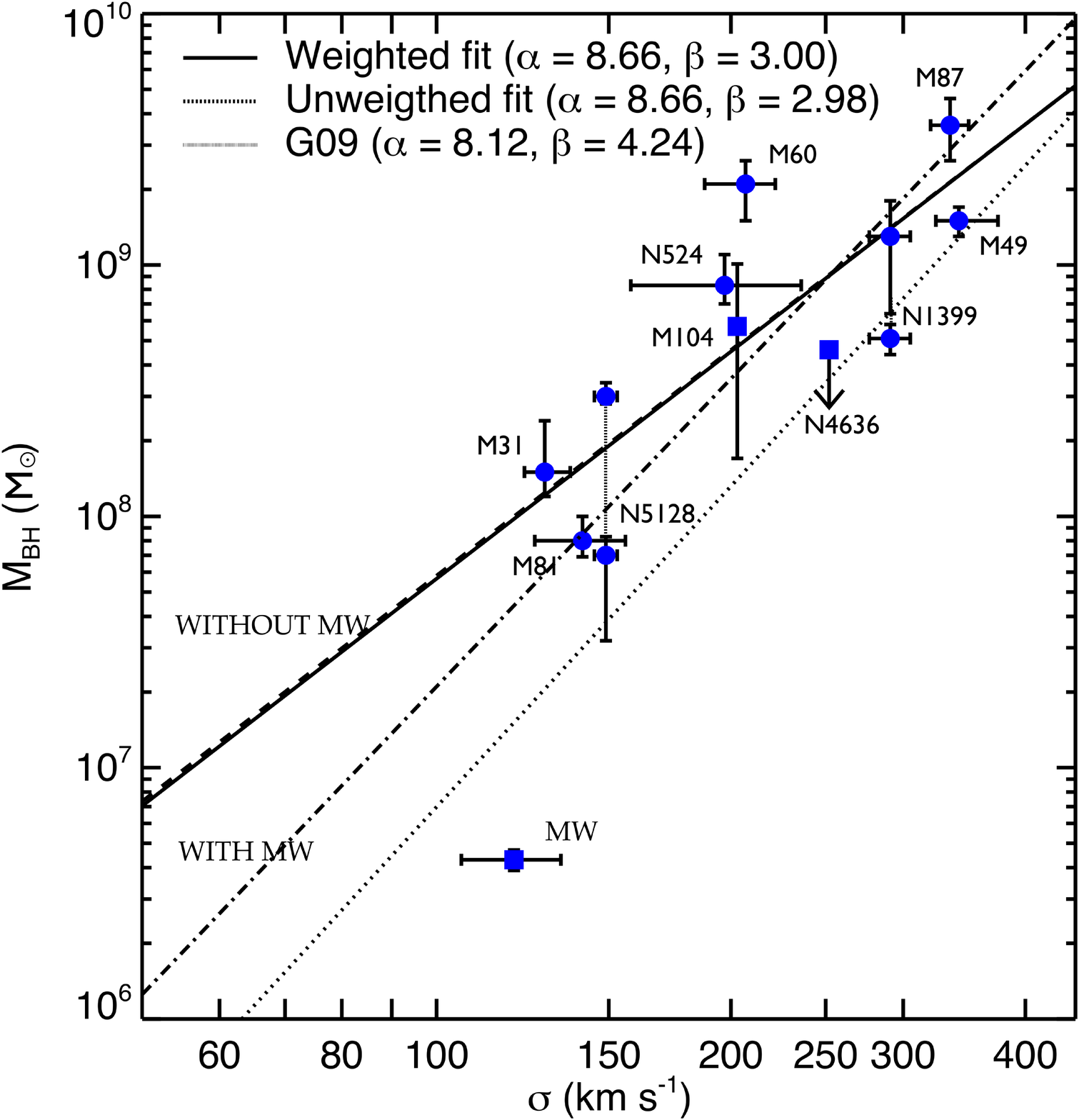}
  \caption{
    Same as Figure \ref{figure:m-sigma red} but considering the velocity dispersion $\sigc^{\rmn{blue}}$
    of the blue, metal-poor globular cluster system. The dash-dotted line is the best-fit obtained by
    including the Milky Way in the fitting procedure.
  }
  \label{figure:m-sigma blue}
\end{figure}

\label{section:m-sigma sub}

GCs are usually divided into two sub-populations with different features : the metal-poor GCs
which are generally older and have a shallower number density profile than the metal-rich GCs
which consist mostly of younger objects associated with the spheroid and lying closer to the centre. 
The two sub-populations are often referred to as the blue and red GCs sub-populations
respectively. Here, we examine separately the $\mbh-\sigc$ relation of these two categories.

Concerning the red GCs, measurements of $\sigc^{\rmn{red}}$ were available for 9 systems in
our sample (including the red sub-population for the Milky Way). The data points as well as the
best-fit relations are shown in Figure \ref{figure:m-sigma red} (the same fitting techniques
detailed in section \ref{section:m-sigma} were used here for the red and blue GCs). The 
relation we find has $\alpha = 8.73 \pm 0.09$ and $\beta = 3.84 \pm 0.52$ with an 
intrinsic scatter $\varepsilon_0 = 0.22$ making it a \emph{tighter} relation
than the $\mbh-\sigma_{sph}$ for the same sample. However, since the Milky Way probes a rather
different region of parameter space and because its black hole mass is known to high 
accuracy compared to other systems, it could potentially introduce a bias in the estimation of the 
parameters. Nonetheless, we found little difference in the resulting parameters even if we removed the Milky Way. We also 
point out that both fitting methods gave almost the same results indicating that the parameters are well-constrained. 
We note that our value of $\beta$ is close to the value given by G09 for 
early-type galaxies ($\beta = 3.86$) and ellipticals ($\beta= 3.96$).

For blue GCs (Figure \ref{figure:m-sigma blue}), we obtain $\alpha = 8.66 \pm 0.12$ and 
$\beta = 3.00 \pm 0.72$ without including the MW. The intrinsic scatter $\varepsilon_0 = 0.33$ is larger 
than the previous one obtained for either the full GC system or for the red GCs and the slope is also 
only marginally consistent with both estimates. Thus, it seems that
SMBHs and metal-poor GCs are only weakly connected. Note that if we include the MW, 
we obtain a higher value for the slope $\beta = 4.07 \pm 0.89$ which is more consistent with G09 and 
with the metal-rich GCs but the scatter is even larger in this case.

Finally, we point out that, according to our best-fit relations for red and blue GCs,
there is a preferred value of $\mbh$ for NGC 1399 and NGC 5128 (Figure \ref{figure:m-sigma red} and 
\ref{figure:m-sigma blue}) which are $\mbh = 3.0 \ \times 10^8 \ M_{\sun}$ and 
$\mbh = 1.3 \ \times 10^9 \ M_{\sun}$ respectively.

%%%%%%%%%%%%%%%%%%%%%%%%%%%%%%%%%%%%%%%%%%%%%%%%%%%%%%%%%%%%%%%%%%%%
\section{Prediction of  $\mbh$ for five galaxies}
%%%%%%%%%%%%%%%%%%%%%%%%%%%%%%%%%%%%%%%%%%%%%%%%%%%%%%%%%%%%%%%%%%%%
\label{section:prediction}

As we have shown, $\sigc^{\rmn{red}}$ seems to be the best proxy for black hole mass for
the sample we considered. Thus, we use the $\mbh-\sigc^{\rmn{red}}$ relation to estimate the black hole mass of galaxies for 
which $\sigc^{\rmn{red}}$ is known, namely for NGC 1407, NGC 3923, NGC 4494, M33 and the LMC (Table \ref{table:masses}).

\begin{itemize}

\item{\emph{M33} -} 
We obtain $\mbh = 1.32 \pm 1.93 \ \times 10^6 M_{\sun}$ which is consistent with M33 having
no central black hole. We remark that \citet{Gebhardt2001} have 
used the $M_{BH}-\sigma_{sph}$ relation ($\alpha = 8.11$ and $\beta = 3.65$) to estimate the black hole mass in M33. 
They found $5.6 \ \times 10^4 M_{\sun}$. \citet{Merritt2001} find $\mbh = 3 \ \times 10^3 M_{\sun}$ 
as an upper limit. They conclude that either M33 does not contain a BH or has an intermediate mass black hole or 
that the $M-\sigma$ relation cannot be extrapolated to such low masses. Using the observed relation between
circular velocity at large radii and black hole mass, \citet{Gebhardt2001} have calculated a mass of 
$\mbh \sim 6 \ \times 10^6 M_{\sun}$. X-ray observations \citep{Foschini2004,Zhang2009,Weng2009} give $\mbh \sim 10 M_{\sun}$. 
Using also X-ray data and a different method, \citet{Dubus2004} find a black hole 
mass in the range $10^5 - 10^6 M_{\sun}$. Further data are needed before a final conclusion can be drawn.

\item{\emph{LMC} -}
The Large Magellanic Cloud, being a satellite galaxy, is clearly different from the 
other systems in the sample used to derive the $\mbh-\sigc$ relation. The centre is not 
well defined and its globular clusters are not old. However, since 
the kinematics of the GC system is available (\citealt{Freeman1993}), we simply mention
that assuming the $\mbh-\sigc$ relation can be applied it predicts $\mbh = 4.1 \pm \ 5.4 \ \times 10^4 M_{\sun}$
which hints towards an intermediate mass black hole but is also consistent with the LMC having no central
black hole.

\end{itemize}

The black hole mass estimates are summarised in Table \ref{table:masses}.

\begin{table}
\centering
\begin{minipage}{70mm}
\caption{
  Estimated black hole mass using the measured
  velocity dispersion of the red GCs sub-population.
}
\begin{tabular}{@{}lccc@{}}
\hline
Galaxy    &  $\sigc^{\rmn{red}}$  &  Ref. & $\mbh$      \\
          &  ($\kms$)            &       & ($M_{\sun}$) \\
\hline 
NGC 1407  &  $243^{+21}_{-16}$    &   1   & $1.12 \pm 0.42 \ \times 10^9$ \\
NGC 3923  &  $200^{+22}_{-22}$    &   2   & $5.31 \pm 2.50 \ \times 10^8$ \\
NGC 4494  &  $92^{+8}_{-21}$      &   3   & $2.69 \pm 2.04 \ \times 10^7$  \\
M33       &  $42^{+18}_{-8}$      &   4   & $1.32 \pm 1.93 \ \times 10^6$  \\
LMC       &  $17$                &   5   & $4.11 \pm 5.36 \ \times 10^4$  \\
\hline
\end{tabular}
\textbf{REFERENCES.} (1) \citealt{Romanowsky2009}; (2) \citealt{Norris2012}; (3) \citealt{Foster2011}; 
(4)\citealt{Schommer1991}; (5)  \citealt{Freeman1993}
\label{table:masses}
\end{minipage}
\end{table}

%%%%%%%%%%%%%%%%%%%%%%%%%%%%%%%%%%%%%%%%%%%%%%%%
 \section{Conclusion }
%%%%%%%%%%%%%%%%%%%%%%%%%%%%%%%%%%%%%%%%%%%%%%%%
\label{section:conclusion}
  
In the present paper, we have shown that the velocity dispersion of 
globular cluster systems projected on the line-of-sight (that is the observed radial velocities) is 
well-correlated with the mass of central black holes, particularly for red (metal rich) 
globular clusters. The slope of the correlation is similar to the one obtained 
by G09 but the normalisation is different, 
meaning that at fixed black hole mass, the velocity dispersion of the globular cluster system is smaller than 
the dispersion of the spheroid.
We also show that the relation for red clusters has the same slope as that of G09 
for early-type galaxies. Thus, it seems that the relation with red globular clusters is an extension
of what happens with $\sigma_{sph}$. For blue globular clusters, the relation is clearly weaker and 
the slope is also lower than the classical $\mbh-\sigma$ relation.
Therefore it is not clear whether a direct connection between dark matter halos, as probed by blue GCs, 
and central black holes really exists or not. We have also used the tight relation with red GCs to estimate 
the black hole mass in NGC 1407, NGC 4494, NGC 3923, M33 and the LMC.

\section*{Acknowledgments}

We wish to thank Karl Gebhardt and Julie Nantais for providing us with unpublished data, 
Roya Mohayaee, Scott Tremaine, Marta Volonteri, William and Gretchen Harris for interesting discussions
and the anonymous referee for relevant comments which helped improving the paper.
Support from ANR OTARIE is acknowledged.

\label{lastpage}

\end{document}